# Patent Novelty Assessment: Accelerating Innovation and Patent Prosecution

1st Kapil Kashyap , 2nd Sean Fargose, 3rd Gandhar Dhonde, 4th Aditya Mishra
*Department of Computer Engineering, Mumbai University Maharashtra, India.*
[1]kapilkashyap3105@gmail.com, fargose.sean2808@gmail.com, gandhardhonde@gmail.com, avm5439@gmail.com

*Abstract*—In the rapidly evolving landscape of technological innovation, safeguarding intellectual property rights through patents is crucial for fostering progress and stimulating research and development investments. This report introduces a groundbreaking Patent Novelty Assessment and Claim Generation System, meticulously crafted to dissect the inventive aspects of intellectual property and simplify access to extensive patent claim data. Addressing a crucial gap in academic institutions, our system provides college students and researchers with an intuitive platform to navigate and grasp the intricacies of patent claims, particularly tailored for the nuances of Chinese patents. Unlike conventional analysis systems, our initiative harnesses a proprietary Chinese API to ensure unparalleled precision and relevance. The primary challenge lies in the complexity of accessing and comprehending diverse patent claims, inhibiting effective innovation upon existing ideas. Our solution aims to overcome these barriers by offering a bespoke approach that seamlessly retrieves comprehensive claim information, finely tuned to the specifics of the Chinese patent landscape. By equipping users with efficient access to comprehensive patent claim information, our transformative platform seeks to ignite informed exploration and innovation in the ever-evolving domain of intellectual property. Its envisioned impact transcends individual colleges, nurturing an environment conducive to research and development while deepening the understanding of patented concepts within the academic community.

*Index Terms*—Patent novelty, prior art search, patent classification, International Patent Classification (IPC), Cooperative Patent Classification (CPC), machine learning, natural language processing, semantic analysis, data security, confidentiality, knowledge management.

## I. Introduction

Determining patent novelty is a critical step in the patent application process, ensuring inventions are truly novel and non-obvious before granting. Traditionally, this novelty assessment relied on manual prior art searches by examiners, which is increasingly challenging due to the exponential growth of patent filings across diverse technologies. In the fast-paced world of technological advancement, protecting intellectual property rights is essential to drive innovation and support creative endeavors. Patents are key to defining the boundaries of groundbreaking inventions and encouraging research, development, and investment. This report explores an advanced system for assessing patent novelty and generating claims, aimed at dissecting the inventive aspects of intellectual property. Automated patent novelty systems, powered by artificial intelligence (AI), machine learning (ML), and natural language processing (NLP), have emerged to address this challenge. These systems streamline novelty evaluations by analyzing proposed inventions against vast global patent databases, using advanced techniques like semantic analysis and similarity scoring. Typically, such systems comprise input modules for technical details, database modules housing patent data (often leveraging APIs like PatentScope), AI modules for analysis/keyword extraction, and output modules suggesting novelty determinations. Effective patent classification schemes like CPC aid targeted searches within specific technical domains. This paper reviews state-of-the-art patent novelty systems, focusing on their components, AI/ML methodologies, and potential to expedite patent examination while fostering innovation. A demonstration using PatentScope's API showcases the system's capabilities in conducting comprehensive prior art searches across global patents.

## II. Literature Survey

The determination of novelty is a critical aspect of the patent application process, as it ensures that an invention is genuinely new and non-obvious compared to existing prior art. In recent years, there has been a growing interest in developing automated patent novelty systems to streamline and enhance this process, leveraging advanced technologies such as artificial intelligence (AI), machine learning (ML), and natural language processing (NLP). Kamateri's study[1] underscores the effectiveness of using ensemble deep learning models, where three bidirectional gated recurrent unit classifiers working concurrently on various patent parts achieved superior accuracy compared to standalone classifiers. This highlights the potential of deep learning ensemble approaches for improved performance in novelty detection. Additionally, a pre-trained BERT model for patent classification[2] demonstrated that relying solely on patent claims can achieve state-of-the-art results, indicating further exploration opportunities in the two-stage framework of pre-training and fine-tuning. Regarding patent eligibility for AI and ML technologies, the European Patent Office (EPO) guidelines generally exclude them unless linked to a technical application, with the revised guidelines refining technical prerequisites without significant alteration[3]. This highlights the importance of considering legal and regulatory aspects when developing patent novelty systems involving AI and ML components. Several studies have explored various factors and approaches for enhancing patent analysis and decision-making. One study[4] developed a predictive model incorporating factors such as technological strength, knowledge accumulation, and technological protection scope to aid in patent transfer decisions. Another study[5] adopted a user-centered approach to integrate patent drawings into the retrieval process, emphasizing the potential for improved image processing methods and increased adoption of patent drawings in search systems. Strategic considerations for patent portfolio management have also been investigated. A study[6] recommended expanding overseas patent layouts and enhancing competition positions and high-strength patents to gain a competitive advantage. Additionally, a study[7] examined the interaction between patent collateral and patent-based measures in the technology market, identifying factors influencing the likelihood and speed of patent collateral. Specific technological domains have also been the focus of patent analysis studies. In the autonomous car industry[8], researchers classified significant subjects encompassing different technologies and study disciplines. Another study[9] analyzed the technological evolution of traffic control systems over the past decade, utilizing patent research to identify advancements and trends while highlighting historical milestones.

Methodological advancements in patent analysis have also been explored. Graph embedding techniques[10] have been proposed to unveil implicit competitive connections between organizations, demonstrating superiority over traditional author-topic modeling strategies in learning competitive linkages within patent networks. Text embedding techniques[11], such as Word2Vec and Doc2Vec, have been employed to assess R and D success, outperforming traditional patent classification code-based approaches. Citation analysis has been a prominent area of research in patent novelty systems. A study[12] introduced the Citation Network and Event Sequence (CINES) model, which utilizes a two-level attention mechanism to encode signals from both citation networks and event sequences, demonstrating superior performance in citation prediction compared to existing techniques. Another study[13] proposed a method combining Main Path Analysis (MPA) with the PageRank algorithm for improved accuracy in identifying main paths and core patents within citation networks. Performance evaluation and optimization of patent analysis systems have also been studied. A study[14] outlined a performance rating model for university patent achievements, incorporating quantitative recursive analysis, fuzzy association rule feature extraction, and fuzzy decision methods, demonstrating enhanced efficiency and accuracy. Additionally, a study[15] examined patent documents using statistical analysis, Python programming, and PostgreSQL DBMS, emphasizing the importance of patent analytics for scientific and commercial entities. Machine learning techniques for merging patent document retrieval results have been explored[16], highlighting the efficacy of random forests and the importance of results merging in distributed information retrieval. Furthermore, a study[17] introduced a novel approach integrating patent citations with text data via a specialized neural network, CCNN, demonstrating its effectiveness in enhancing patent similarity comparison, clustering, and map generation. Finally, the importance of early identification of disruptive emerging technologies has been recognized[18], proposing a research framework that supplements traditional academic papers and patent data with web news data to leverage social awareness for identifying development trends. This literature review highlights the diverse range of research efforts aimed at developing and improving patent novelty systems, encompassing various technological domains, methodological approaches, and strategic considerations. As the volume and complexity of patent data continue to grow, further advancements in this field will be crucial for promoting innovation, protecting intellectual property rights, and streamlining patent examination processes.

## III. Proposed Design

The project envisions a user-friendly website for students and academics to access patented concepts. The website will employ backend APIs to fetch and compile patent claim data, including a Chinese API for comprehensive coverage of Chinese patents. Users can navigate an organized interface to retrieve detailed insights into specific patent claims. This design emphasizes accessibility, precision, and relevance for innovation in academia. The fetched API data can be stored in a proprietary database and updated annually through web scraping to maintain a robust patent information repository.

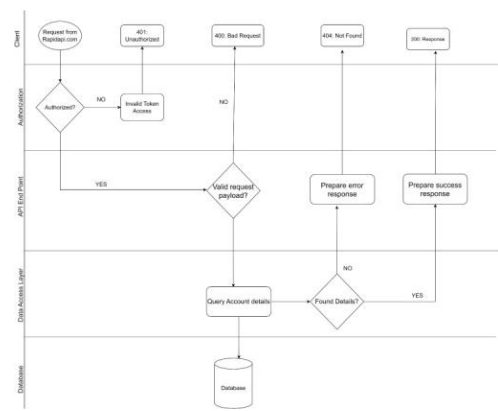

Fig 2.1
The API Architecture

Figure 2.1 depicts the flow of handling requests to the Global Patent API system. The client, identified as Rapidapi.com, initiates a request. The first step is to authenticate the request by validating the provided Access Token. If the Access Token is invalid, a 401 Unauthorized response is returned. With a valid Access Token, the request payload is validated to ensure the format and data conform to requirements. An invalid payload results in a 400 Bad Request response. The system queries the Account Details from a database (DB) with a valid request payload. The user login and registration data is stored in a PostgreSQL database managed by pgAdmin. Based on the retrieved data, the system determines if the requested details are found in the database. If the requested details are not available, a 404 Not Found response is sent back to the client. However, if the requested details are found, a 200 Response containing the requested data is prepared and returned to the client, completing the request cycle successfully. This architecture demonstrates a secure API system that authenticates users, validates incoming requests, retrieves user data from a PostgreSQL database, and responds appropriately based on the request and data availability. It follows industry-standard practices for building secure and robust API's that handle user authentication and data retrieval from a database.

## IV. IMPLEMENTATION

In this initial implementation of the GUI, users are presented with a streamlined interface that harnesses data fetched through the PatentScope API, empowering them to efficiently explore and analyze patents of interest. With intuitive navigation and comprehensive search capabilities, this GUI provides a user-centric platform for accessing pertinent patent information.

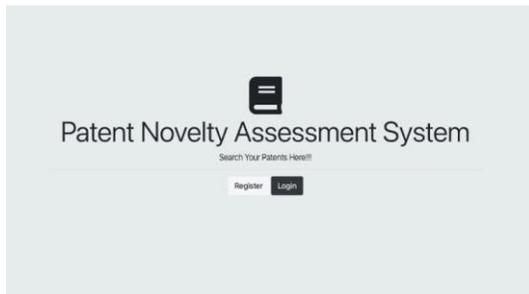

Fig. 4.1 Home page

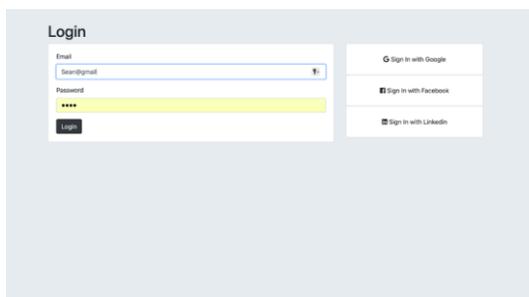

Fig. 4.2 Login page

Moreover, the integration of the PatentScope API not only facilitates the retrieval of patent data but also lays the foundation for a robust patent novelty system. By leveraging advanced algorithms and classification schemes, this system will enable users to assess the novelty of their inventions and identify relevant prior art with greater accuracy and efficiency. Through seamless integration with the GUI, the patent novelty system will enhance users' ability to make informed decisions during the patent application process, ultimately fostering a culture of innovation and intellectual property protection. As this GUI evolves, its integration with the patent novelty system will further empower users to navigate the complex landscape of intellectual property with confidence and precision.

## V. RESULTS

```
Application Date Of Patent : 2023-04-21
Application Number Of Patent : CN202320905257.6
ID Of Patent : CN219938033U
Result No. 2
Application Date Of Patent : 2016-01-14
Application Number Of Patent : CN201610022518.4
ID Of Patent : CN105499068A
Result No. 3
Application Date Of Patent : 2016-01-14
Application Number Of Patent : CN201610022518.4
ID Of Patent : CN105499068B
Result No. 4
Application Date Of Patent : 2015-04-30
Application Number Of Patent : CN201510215461.5
ID Of Patent : CN104839950A
Result No. 5
Application Date Of Patent : 2018-06-12
Application Number Of Patent : CN201820902075.2
ID Of Patent : CN208479216U
Result No. 6
Application Date Of Patent : 2015-04-30
Application Number Of Patent : CN201510215461.5
ID Of Patent : CN104839950B
Result No. 7
Application Date Of Patent : 2021-06-03
```

Fig. 5.1 shows the results obtained for all patents under the name Apple Watch

By utilizing the Global Patents API, we have gathered an extensive dataset that includes essential information on patent applications. This dataset features the filing dates, the total count of patents, and unique patent identifiers. Additionally, it includes the titles of the patents, enhancing our understanding of the intellectual property protected. This comprehensive data set supports thorough analyses and precise tracking of patent activities, equipping users to make informed decisions and sustain a competitive edge in their fields. The acquired information is crucial for those involved in patent management, legal evaluations, and intellectual property studies. Access to such detailed patent information is vital for anyone looking to effectively manage and understand the complexities of intellectual property rights.

## VI. FUTURE SCOPE

The future road map involves broadening the system's scope to encompass patent information from diverse languages and regions, extending beyond its current focus on Chinese patents. Advanced natural language processing (NLP) and machine learning (ML) techniques will be integrated to enhance the accuracy and efficiency of patent claim analysis. Additionally, the system will be equipped with automated capabilities to generate patent claims, streamlining the drafting process. Seamless integration with existing patent drafting, management systems, and research/innovation platforms is planned to provide a cohesive ecosystem. To foster collaboration and knowledge sharing, a dedicated platform will be developed, allowing stakeholders to contribute insights, discuss patent claims, and collectively enrich the knowledge base. Furthermore, the system will incorporate patent landscape analysis and visualization tools, offering insights into emerging technologies, key players, and trends within specific domains. Ensuring the system's relevance and adaptability is a key priority. Continuous improvement efforts will be undertaken to align with evolving patent laws, regulations, and user feedback. This comprehensive approach leverages cutting-edge AI/ML models, optimizes patent drafting workflows, cultivates a community-driven knowledge repository, provides valuable technological insights, and maintains a future-ready stance in the dynamic intellectual property landscape.

## VII. Conclusion

In conclusion, our envisioned project marks a significant leap forward in addressing the complexities surrounding intellectual property, particularly within the realm of Chinese patents, for college students and academic institutions. Through the development of a user-friendly website leveraging diverse APIs, we aim to streamline the often daunting process of accessing patent claims. By harnessing technology and specialized APIs, our project not only simplifies access to comprehensive patent information but also fosters a nuanced understanding of the unique landscape of Chinese patents.

This initiative holds tremendous potential in nurturing a culture of innovation within academia. It empowers students and institutions to delve into, comprehend, and build upon existing patented concepts. Our holistic approach is finely attuned to the evolving needs of both academic and innovative communities, promising to advance knowledge and creativity within the intellectual property sphere.

As we embark on the journey to develop this trans-formative platform, we anticipate its impact will extend far beyond academic boundaries. By facilitating informed exploration and innovation, our project stands poised to catalyze positive change within the ever-evolving domain of intellectual property, driving forward progress and fostering a vibrant culture of innovation worldwide.

## References

[1] Eleni Kamateri, Vasileios Stamatis, Konstantinos Diamantaras, and Michail Salampasis. 2022. Automated Single-Label Patent Classification using Ensemble Classifiers. In Proceedings of the 2022 14th International Conference on Machine Learning and Computing (ICMLC '22). Association for Computing Machinery, New York, NY, USA, 324–330. https://doi.org/10.1145/3529836.3529849

[2] Aaron Chu, Shigeyuki Sakurai, and Alfonso F. Cardenas. 2008. Automatic detection of treatment relationships for patent retrieval. In Proceedings of the 1st ACM workshop on Patent information retrieval (PaIR '08). Association for Computing Machinery, New York, NY, USA, 9–14. https://doi.org/10.1145/1458572.1458575

[3] L. Wang and S. Hu, "Patent Protection for Artificial Intelligence in Europe," 2020 International Conference on Intelligent Transportation, Big Data and Smart City (ICITBS), Vientiane, Laos, 2020, pp. 591-594, doi: 10.1109/ICITBS49701.2020.00130.

[4] Kim and Y. Geum, "Predicting Patent Transactions Using Patent-Based Machine Learning Techniques," in IEEE Access, vol. 8, pp. 188833-188843, 2020, doi: 10.1109/ACCESS.2020.3030960.

[5] Wiebke Thode. 2021. Integration of Images into the Patent Retrieval Process. In Proceedings of the 2021 Conference on Human Information Interaction and Retrieval (CHIIR '21). Association for Computing Machinery, New York, NY, USA, 351–354. https://doi.org/10.1145/3406522.3446006

[6] X. Yang and X. Yu, "Identifying Patent Risks in Technological Competition: A Patent Analysis of Artificial Intelligence Industry," 2019 8th International Conference on Industrial Technology and Management (ICITM). 2019. doi:10.1109/icitm.2019.8710719.

[7] H. -N. Su, "How Do Patent-Based Measures Inform Patent Collateral? A Holistic Analysis on All USPTO Patents Between 1986 and 2016," in IEEE Transactions on Engineering Management, vol. 69, no. 6, pp. 3265-3275, Dec. 2022, doi: 10.1109/TEM.2020.3038855.

[8] J. Ko and J. Lee, "Discovering Research Areas from Patents: A Case Study in Autonomous Vehicles Industry," 2021 IEEE International Conference on Big Data and Smart Computing (BigComp), Jeju Island, Korea (South), 2021, pp. 203-209, doi: 10.1109/BigComp51126.2021.00046.

[9] A. Durmusoglu and Z. D. Unutmaz Durmusoglu, "Traffic Control System Technologies for Road Vehicles: A Patent Analysis," in IEEE Intelligent Transportation Systems Magazine, vol. 13, no. 1, pp. 31-41, Spring 2021, doi: 10.1109/MITS.2020.3037319.

[10] .Wang, Yunli and Richard, Rene and McDonald, Daniel. (2020). Competitive Analysis with Graph Embedding on Patent Networks. 10-19. 10.1109/CBI49978.2020.00009.

[11] .Ha, Taehyun, and Jae-Min Lee. 2023. "Examine the Effectiveness of Patent Embedding-Based Company Comparison Method." IEEE Access. https://doi.org/10.1109/access.2023.3251664.

[12] Fang He, Wang-Chien Lee, Tao-Yang Fu, and Zhen Lei. 2021. CINES: Explore Citation Network and Event Sequences for Citation Forecasting. In Proceedings of the 44th International ACM SIGIR Conference on Research and Development in Information Retrieval (SIGIR '21). Association for Computing Machinery, New York, NY, USA, 798–807. https://doi.org/10.1145/3404835.3462903

[13] Lu, Z., Ma, Y., Song, L. (2021). Patent Citation Network Analysis Based on Improved Main Path Analysis: Mapping Key Technology Trajectory. In: Sun, X., Zhang, X., Xia, Z., Bertino, E. (eds) Advances in Artificial Intelligence and Security. ICAIS 2021. Communications in Computer and Information Science, vol 1423. Springer, Cham.

[14] Zhang Liang and You Anni. 2021. Design of Performance Evaluation System for Transformation of Patent Achievements in Colleges and Universities Based on AHP. In 2021, the 2nd International Conference on Computers, Information Processing and Advanced Education (CIPAE 2021). Association for Computing Machinery, New York, NY, USA, 1070–1076. https://doi.org/10.1145/3456887.3457463.

[15] A. Glushko and D. S. Silnov, "Analytics in the Field of Patent Documents," 2021 IEEE Conference of Russian Young Researchers in Electrical and Electronic Engineering (ElConRus), St. Petersburg, Moscow, Russia, 2021, pp. 364-368, doi: 10.1109/ElConRus51938.2021.9396375.

[16] Vasileios Stamatis and Michail Salampasis. 2021. Results Merging in the Patent Domain. In Proceedings of the 24th Pan-Hellenic Conference on Informatics (PCI '20). Association for Computing Machinery, New York, NY, USA, 229–232. https://doi.org/10.1145/3437120.3437313

[17] M. Li and H. Li, "Application of Deep Convolutional Neural Network Under Region Proposal Network in Patent Graphic Recognition and Retrieval," in IEEE Access, vol. 10, pp. 37829-37838, 2022, doi: 10.1109/ACCESS.2021.3088757.

[18] J. Qi, L. Lei, K. Zheng, and X. Wang, "Patent Analytic Citation-Based VSM: Challenges and Applications," in IEEE Access, vol. 8, pp. 17464-17476, 2020, doi: 10.1109/ACCESS.2020.2967817.

[19] X. Li, Q. Xie, and L. Huang, "Identifying the Development Trends of Emerging Technologies Using Patent Analysis and Web News Data Mining: The Case of Perovskite Solar Cell Technology," in IEEE Transactions on Engineering Management, vol. 69, no. 6, pp. 2603-2618, Dec. 2022, doi: 10.1109/TEM.2019.2949124.

[20] .G. Yue, J. Liu, Y. Hou, and Q. Zhang, "A Novel Patent Knowledge Extraction Method for Innovative Design," IEEE Access, vol. 11. pp. 2182–2198, 2023. doi: 10.1109/access.2022.3229490.